\documentclass[paper]{geophysics}

\usepackage{indentfirst}
\usepackage{tabularx}
\usepackage{xcolor}
\usepackage{ulem}
\begin{document}

\title{Complete identification of complex salt-geometries from inaccurate migrated subsurface offset gathers using Deep Learning}
\righthead{Deep learning salt detection} 

\author{Ana Paula O. Muller\footnotemark[1] \footnotemark[2] \footnotemark[6],
        Jessé C. Costa\footnotemark[3] \footnotemark[4],
        Clecio R. Bom\footnotemark[1] \footnotemark[5], 
        Elisangela L. Faria\footnotemark[1],
        Matheus Klatt\footnotemark[1], 
        Gabriel Teixeira\footnotemark[1],
        Marcelo P. de Albuquerque\footnotemark[1], 
        Marcio P. de Albuquerque\footnotemark[1]}

\maketitle
\begin{abstract}
Delimiting salt inclusions from migrated images is a time-consuming activity that relies on highly human-curated analysis and is subject to interpretation errors or limitations of the methods available. We propose to use migrated images produced from an inaccurate velocity model (with a reasonable approximation of sediment velocity, but without salt inclusions) to predict the correct salt inclusions shape using a Convolutional Neural Network (CNN). Our approach relies on subsurface Common Image Gathers to focus the sediments' reflections around the zero offset and to spread the energy of salt reflections over large offsets. Using synthetic data, we trained a U-Net to use common-offset subsurface images as input channels and the correct salt-masks as output of a semantic segmentation problem. The network learned to predict the salt inclusions masks with high accuracy; moreover, it also performed well when applied to synthetic benchmark data sets that were not previously introduced.
Our training process tuned the U-Net to successfully learn the shape of complex salt bodies from partially focused subsurface offset images.
\end{abstract}

\section{Introduction}


Velocity model building (VMB) is essential to make accurate subsurface images, especially in regions with high contrast velocities and complex structures. The conventional model building process uses seismic ray-based tomography \cite[]{tomo,RAYTOMO} and Full Waveform Inversion (FWI) \cite[]{tarantola,pratt} to determine the velocity model. Each iteration of Tomography or FWI requires a high level of regularization and a good input model to avoid local minima or cycle skipping. The result is often bounded to the maximal resolution power of the method, which often does not resolve the geological complexity of the area. One good example of the limitation of conventional inverse methods is the inclusion of salt bodies in the velocity model, a complex non-automated method that poses a high cost and is subject to uncertainties.

Salt inclusion is critical during the VMB. Whence a mistake in salt geometry makes the image below the salt unfocused or distorted \cite[]{gardenbanks}, generating a wrong structure of the subsurface.  Therefore, it can lead to economic consequences, especially in petroleum provinces where the reservoirs are below complex salt structures. 
Salt presents a great diversity of possible geometries \cite[]{HUDEC20071} and has a high velocity contrast with the enclosing sediments. In complex areas, the definition of the salt geometry is estimated in an iterative process called salt flood \cite[]{saltflood}. Salt flood iterations require massive interpretation, geological knowledge of the sedimentary basin, and scenarios testing \cite[]{saltscenarios01,saltscenarios02}. Moreover, each iteration requires a lot of computational effort for migrating the data and observing the respective response of the interpreted salt geometry in the quality of the image generated. Some works try to overcome such difficulties by focusing on FWI to correct the badly interpreted salt \cite[]{fwisalt01,fwisalt02,FWIvariance}. However, it still requires data acquired with long offset, wide-azimuth, and low frequency, or the imposition of geological constraints over the method \cite[]{fwic}.

The problem of identifying the presence of salt in seismic images was previously studied using Deep Learning (DL) \cite[]{SALTdetection,salt:SEG}. Nonetheless, this salt-segmentation is made over a well-focused migrated image. The focalization and accuracy of a seismic image are directly dependent on an accurate velocity model used in migration, where the model must include a reasonable estimation of the salt structures above the interpretation zone. Therefore, the available algorithms for salt-segmentation do not eliminate the salt flood steps during the VMB. These segmentation techniques have a high value for interpretation from seismic images and for refinement of salt structures. However, these structures must be somehow present in the velocity model used to migrate the seismic data.

Another process that attracts attention in applying DL is velocity estimation from seismic data \cite[]{ARAYAPOLO1,ovchar1}. Many recent works are using DL aiming to make the complete prediction of velocity models, being particularly successful in predicting the salt bodies embedded in such models, even those with very complex geometries \cite[]{ARAYAPOLO2,YANGBASE,shucai20}. These works use the raw seismic shot data to train the network to fully predict the correct velocity model that generated those seismic shots.  Their results are very encouraging; however, they can not be easily extended to current seismic field data. The first reason is the size of real seismic acquisitions. DL needs to hold the training data in the device memory, and real seismic shots are much larger than the most modern device could support, and even if they could be supported, the number of convolution operations would be prohibitive. Another difficulty is the irregularity of seismic shots, and the geometric non-correspondence of this data with the seismic images or the velocity models, i.e., source and receivers vary their positions during the acquisition process. Concurrent to our work, \cite{Fomel2021}, have independently proposed a new approach that bypasses some problems mentioned here. Their work uses the common image gathers (CIG) in the angle domain and presents good results in recovering with DL structurally simple models from synthetic data sets.

Similarly, in this work, we propose to reduce the input data to the imaging domain to deal with more realistic seismic acquisition geometries, making a hybrid approach of the conventional VMB workflow and DL velocity estimation from seismic data. 
Furthermore, we use DL only in the most challenging step of VMB flow: the inclusion of salt structures. Instead of using raw shots as input to the network, we migrated these shots with a reasonable sediment velocity model, without any salt inclusions. Our migration process uses an extended imaging condition, to generate a set of images with different subsurface offsets, which partially focus incorrectly migrated salt reflections. These images are the input channels for a DL algorithm to solve a segmentation problem, which has as output the salt mask. Our approach accelerates the complex workflow for salt estimation, which conventionally requires  multiple migration and interpretation steps. It differs from previous works on salt segmentation over stacked migrated images \cite[]{SALTdetection,SALTUNET,salt:SEG}, which fits on an advanced stage of VMB flow when there is a reasonable estimate for the salt inclusion. 
The trained U-Net returned accurate predictions over the test set and also over benchmark salt models, indicating that this approach has robust features when compared with traditional methods.
Since it requires only focusing information, it is less sensitive to the velocity model of sediment, i.e., the training and test sets do not need to present the same model distribution. Besides, the evaluation of the loss function on the migrated domain attenuates the effects of coherent and incoherent noise. 
Moreover, even when the results present high loss measures, the estimated salt masks are not drastically off, and can still be used to improve interpretation.

\section{RTM with extended imaging condition}

Reverse Time Migration (RTM) is one of the best algorithms for
seismic imaging data in areas of high complexity \cite[]{etgen2019}. It can handle high structural dip and velocity contrast, conditions common in salt basins. The inputs for RTM are the seismic data shots and the parameters models of the subsurface. For simplicity and to validate the idea, we will study only isotropic 2-D models in this work. 

RTM uses the full wave equation to back propagate the fields of the receivers $W_{r}(\mathbf{x},t)$  and to forward propagate the field of the source $W_{s}(\mathbf{x},t)$ into the earth, where $\mathbf{x}=\{x,z\}$ for our 2-D case. This propagation is made individually for each shot, applying an imaging condition at each time step of the wave equation solution. In this work, we will use a cross-correlation extended imaging condition \cite[]{extIC}, whose generalized form can be written as:
\begin{equation}
R(\mathbf{x},\mathbf{\lambda},\tau)=\sum_{shots} W_{s}(\mathbf{x}-\mathbf{\lambda},t-\tau)W_{r}(\mathbf{x}+\mathbf{\lambda},t+\tau).
\label{eq:ic}
\end{equation}

\begin{figure}
  \centering
  \includegraphics[width=0.20\textwidth]{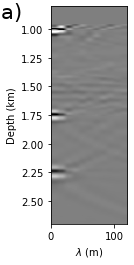}
  \includegraphics[width=0.20\textwidth]{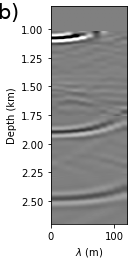}
  \includegraphics[width=0.20\textwidth]{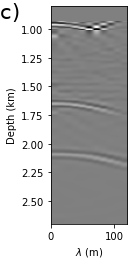}
  \caption{Space-lag common-image-gather migrated with (a) correct velocity, (b) low velocity, and (c) high velocity.}  
  \label{fig:CIG}
\end{figure}

The correlation between source and receiver wave-fields is also evaluated around the zero lag in the extended imaging condition.
Correlations at non-zero lags indicate that the velocity model is incorrect. Equation \ref{eq:ic} opens the possibility to vary the time and the space lag. For simplicity, we will focus only on horizontal spatial extension, thus keeping $\tau=0$, and applying $\mathbf{\lambda}$ only over the $x$ direction. $\lambda$ is known as the subsurface offset, and it has one interesting property that differs from the well-known surface offset, which was decisive in choosing this representation in our study. 

\begin{figure}
  \centering
  \includegraphics[width=0.41\textwidth]{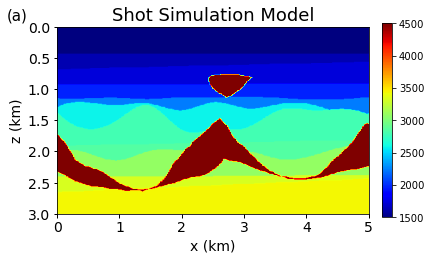}
  \includegraphics[width=0.41\textwidth]{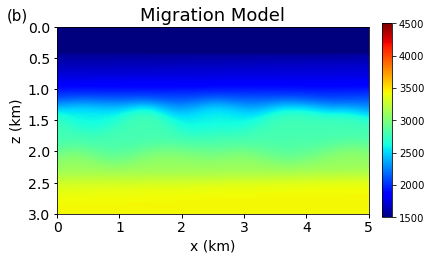}
  \includegraphics[width=0.32\textwidth]{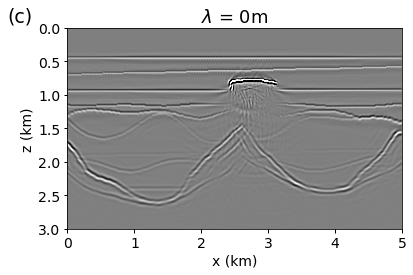}
  \includegraphics[width=0.32\textwidth]{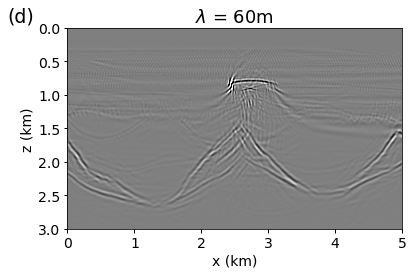}
  \includegraphics[width=0.32\textwidth]{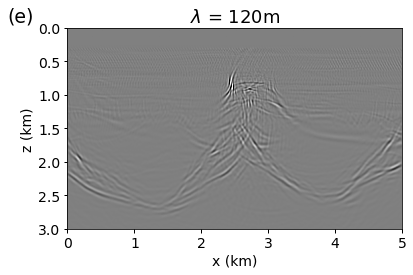}
  \caption{Example of one model used to simulate the shots (a) and the correspondent smoothed version without salt inclusion used for migration (b). The second line shows the effect of the lack of salt in the migration model along the common subsurface offset panels ($\lambda$), where the salt reflections dominate the large offsets due to the velocity error. The acquisition parameters used to generate this result are described in Section Modeling Seismic Acquisition.}  
  \label{fig:channels}
\end{figure}

A gather presenting the subsurface offset, migrated with the correct velocity model, concentrates the reflections' energy in $\lambda=0$. The focus of energy in zero subsurface offset occurs because the source and receiver fields correlate better at the correct position when using the right model. To illustrate this property, in Figure \ref{fig:CIG}(a), we show a gather migrated with the right velocity model. When the velocity used in RTM is not the correct one, the unfocused energy due to the error in the model spreads through the far offsets. In Figure \ref{fig:CIG}(b), we can see a gather migrated with low velocity, and in Figure \ref{fig:CIG}(c), the case with a high velocity. Few studies make use directly of the subsurface offset gathers \cite[]{TWI:IMAGE}; usually, the gathers are converted to the angle domain, and these angle gathers continue the flow of VMB \cite[]{angle}.

In this work, we propose using the property of the spreading energy along $\lambda$ to detect the high contrast between the velocities of salt and sediment. Consider that we migrate the data with a reasonable velocity model of the sediment, without any salt body included, then majorly the salt energy will spread through the non-zero offset.  

In Figure \ref{fig:channels} we illustrate an example of inaccurately migrated salt. On the first line, we show the model used to generate synthetic shots (a) and the model used in RTM migration (b). The model used in RTM is a smoothed version of sediment without salt inclusion. On the second line, we show the common subsurface offset panels, with $\lambda$ = 0 (c), $\lambda$ = 60~m (d) and $\lambda$ = 120~m (e). As can be seen in Figure \ref{fig:channels}, the energy of the unfocused salt will dominate the large offsets. Of course, the shape of salt bodies will be completely distorted and compressed due to the error in the migration velocity, and these distortions vary along the $\lambda$. Indeed to infer the salt geometry using only the inaccurate migrated images presented in Figure \ref{fig:channels} is a task which requires mapping the non-linear relations between input and output data. We will show in the following sections that a CNN can map the images over the subsurface offsets to the right salt-body shape.

\section{Data set definition}

We generated a set of 1505 synthetics data sets based on marine 2-D models, to validate the idea of using subsurface offset as input of a CNN to determine salt geometry. The models have an area of 3~km of depth and 5~km of lateral extension. The following sections describe the complete process of data set definition, from choosing the appropriated salt geometries, velocity model definition, modeling of the shots, and finally migration with the extended imaging condition.

\subsection{Salt geometries and the velocity models}

In order to get good responses from the network when presented to geological salt geometries, we obtained the salt geometries used in this work from real models built with tomography and FWI in the past decade in regions with complex salt structures and high exploration interest. We extracted salt masks from these models as 2-D slices with a minimal increment of 0.5~km to achieve a wide variety of structures. Furthermore, we distorted the proportion between vertical and horizontal sizes to fit the salt body in the size of our experiments(3~km x 5~km). With this strategy, we generated a total of 215 salt masks.

\begin{figure}
  \centering
  \includegraphics[height=0.21\textwidth]{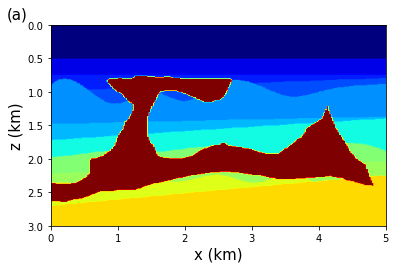}
  \includegraphics[height=0.21\textwidth]{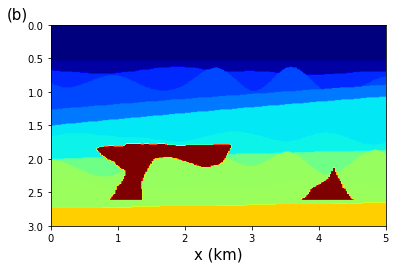}
  \includegraphics[height=0.21\textwidth]{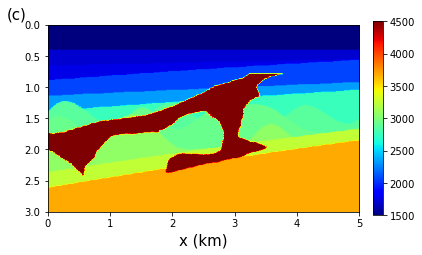} \\
  \caption{Three examples of models from the set of $1505$ synthetic models used in this work. Figure (a) includes a geological salt mask, and (b) and (c) use distortions of this mask.}  
  \label{fig:exmode}
\end{figure}

Besides, we applied distortions in these geological salt masks to increase the size of the original data set and the variability of geometries presented to the network. The distortions applied consist of randomly combined operations, like vertical and horizontal shifts, zoom-in or zoom-out, and rotations. First, we cut our mask in the top and bottom in 400~m to make the distortions, working over a 2.2~km of depth mask. These cuts were made to avoid salt extrusions and ensure that there will be at least one sediment layer below the bottom of the deepest salt in each model. Then, we included top and bottom layers of 400~m without salt to return the masks generated to the interest size. 

There were generated six extra masks from the original one, making a total of 1505 salt masks to define the training, validation, and test set. With this process, we hope to increase our network's ability to generalize to any salt geometry. 
We included each salt mask in different sediment velocity models. The sediment models
are randomly generated obeying some rules. The first layer has a velocity of 1500~m/s, and the last one has a velocity of 3500~m/s. Each model has eleven velocity layers with random shapes and depths. The salt body is then inserted with a velocity of 4500~m/s. In Figure \ref{fig:exmode} we show three examples of models from our data set. Figure \ref{fig:exmode}(a) has an original salt mask, extracted from a real model, Figure~\ref{fig:exmode}(b) and Figure~\ref{fig:exmode}(c) have a salt obtained by applying distortions over the original mask.

\subsection{Modeling seismic acquisition}
\label{sec:modeling}

For each model, we simulated a synthetic seismic acquisition, using a finite-difference wave propagator, isotropic, acoustic, with second-order in time and eight-order in space (with spatial sampling equals to 10~m), and CPML boundaries \cite[]{MartinPML,KOMASTISCHPML}. To simulate the source, we used a Ricker wavelet with a peak frequency of 20Hz. The shots are recorded by 601 receivers with a distance of 10~m between them, in a split spread acquisition with the maximum offset relative to the source of 3~km. There were recorded 125 shots, the first shot occurs at the beginning of the interest region and the last shot at the end, with 40~m of interval between them. Receivers record at a rate of 4~ms for 3.5~s. We applied the mute of direct wave in shots for migration. The models were laterally extrapolated to allow the full shot covering inside the interest area. 

We chose these acquisition parameters to ensure a minimum quality level for the migrated images since the quality of the input images affects the network performance. However, due to the computational cost of simulations, we did not try to investigate the optimum parameters for network performance.

\subsection{Migrated images for training}

As mentioned in the previous sections, we migrated our shots with an RTM cross-correlation extended imaging condition defined in equation \ref{eq:ic}. The subsurface offset $\lambda$ starts at 0m , which is equivalent to the non-extended condition, with an increment of 20~m.

Our main objective in introducing the use of a migration method was to preserve the CNN's ability to predict the salt geometries, while reducing the amount of information to input into the network.
We chose RTM due to its capacity to image complex structures, but other migration methods could be tested in future extensions of this work.

When we propose to migrate the shots, it is necessary to input also a velocity model in depth. Thus we gain in reducing the size of inputs, but we lose with the requirement of such model.
Considering that the conventional salt flood process occurs after the initial definition of the sediment model, it is also reasonable to include our method in this step of VMB flow. The sediment model for a salt flood can be a low-frequency model, like the ones obtained with interactions of tomography and does not need to honor all the details of the layers. To emulate a reasonable sediment velocity model for salt inclusion, we smoothed the slowness using a Gaussian filter with $\sigma=12$. Smoothing the slowness preserves most reflections' travel time but leaves some errors in the fine-scale.

\section{Deep Learning pipeline for salt mask prediction}

The presented deep learning pipeline for salt mask prediction is based on an U-Net architecture. The U-Net \cite[]{UNET} is a type of fully convolutional network \cite[]{LONGSHELHAMER} model widely used in tasks related to semantic segmentation \cite[e.g.,][]{SEGNET}. The model's encoder-decoder structure is ideal for tasks where network inputs and outputs must have the same spatial dimensions. At least a couple of DL applications in seismic analysis problems have used the U-Net topology \cite[e.g.,][]{SALTUNET,DENOISE}, despite not being strictly related to semantic segmentation tasks. This is possible since these problems can often be formulated with 2D, or 3D, inputs and outputs of equal spatial dimensions, which works well with networks of this type.

Here we pose the problem of mapping salt inclusions from migrated sections as one of binary mask prediction. The U-Net receives a [256, 512,  $N_{\lambda}$] volume as input, where $N_{\lambda}$ is the number of subsurface offsets panels, and produces as output a [256,512,1] probability map with high peaks over zero and one, and with the residuals samples concentrated around these peaks. In this case, the threshold to show the predicted mask is naturally at the value of 0.5.
With respect to the original design \cite[]{UNET}, we have made only two changes to the architecture: 1) our convolutional block, in both encoder and decoder sections, contains a batch normalization layer between each pair of convolutional and activation layers, which helps to stabilize the gradient flow through the network \cite[]{BNORM}; 2) our network starts with 16 feature maps in the first stage of the encoder, which significantly reduces the number of network parameters to be trained.

\begin{figure}
  \centering
  \includegraphics[width=0.90\textwidth]{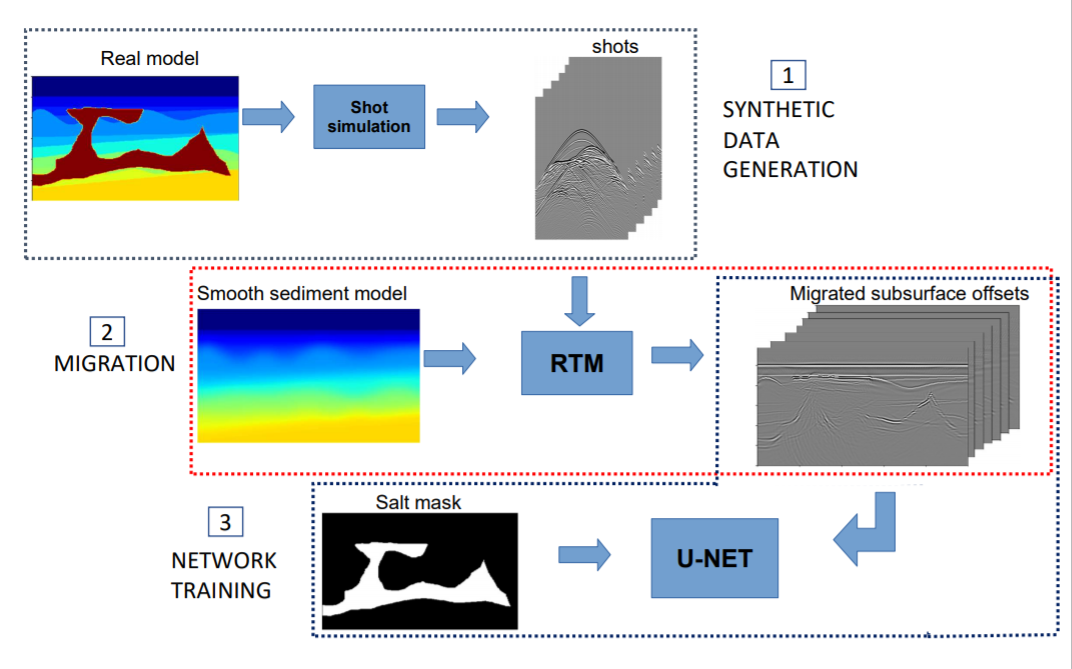}\\
  \includegraphics[width=0.90\textwidth]{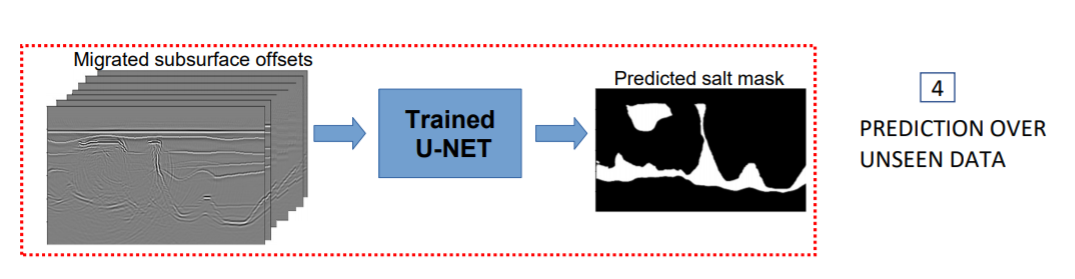}
  \caption{Schematic of our flow. We divided the flow into four steps 1) the generation of the training data set, 2) migration of shots to reduce the information and make the data domains compatible for the network, 3) network training, and finally 4) the salt mask prediction over unseen data.}  
  \label{fig:flow}
\end{figure}

The four stages encoder section of the network gradually downsamples the data at each stage and learns to extract relevant information to the task at different resolution scales of the data. Downsampling occurs only in the first three encoder stages, where it is performed by max-pooling layers after processing through the convolutional block.
The decoder section gradually upsamples the feature map back to the original spatial dimensions. It also has four stages. At each stage the data is first upsampled with a transpose convolutional layer, then concatenated with the feature map coming from the skip connection of the corresponding encoding stage, before going through the convolutional block. 
After the last upsampling stage in the decoder, a final convolutional layer is applied to the data to reduce the number of channels to one, so it matches the single channel of the salt mask labels. Then a sigmoid activation function is computed over this single panel ([256, 512, 1]), which represents the network's predictions. 

The total number of trainable parameters is equal to 2,162,225 when using an input with $N_{\lambda}=7$. 
We apply the Adam optimizer \cite[]{ADAM} to update the parameters with a learning rate equal to 6e-4. Our model is built with TensorFlow \cite[]{TF}. All training processes are carried out on one NVIDIA V100 with 32Gb, and one training process of 60 epochs takes approximately ten minutes.

In Figure \ref{fig:flow} we summarized our flow from the data generation process to the predictions over unseen masks. Step 1 is usually adopted in the literature of VMB with DL, which consists in the generation of synthetic seismic acquisitions from a known model. This model is used to generate the labels for the network training.. The main novelties in our flow are in  step 2), where we changed the input data representation to migrated subsurface offsets panels, and in steps 3) and 4), with training and prediction of the salt mask over this new input data set. It is essential to clarify that the prediction step does not require any previous salt geometry information.
The RTM common image gathers reduced the original seismic data size by sixty times, bringing the network input to a more affordable size. Moreover, the network input and output are in the same domain and have the same size; such features simplify the DL approach since the U-Net does not need to perform a domain transformation.

\subsection{Training methodology}

We trained our network with $N_{\lambda}$~=~7, i.e. with seven subsurface offsets panels, which varies from $\lambda$~=~0~m to $\lambda$~=~120~m with 20~m of increment. These panels work like channels of the input image for the CNN. The original size of salt masks and migrated panels is 300 samples in depth and 500 samples laterally. In order to fit the input and output images to the proposed U-Net design, we cut the images in depth to 256 and pad 6 samples at each lateral to have images with 256x512 samples. In addition, since the models have a minimal water layer, we cut the top portion of the images.
We also clipped the amplitude of each subsurface offset panel in percentile 96 and then normalized the amplitude of each panel between 0 and 1. 

We  used the $1505$ synthetic models previously described. This data is split into three distinct sets: training, validation, and test. The training set is the one that contributes to updating the network's weights. The validation set is used to control the metrics and loss during the training process to avoid overfitting. Finally, the test set is used to make the network predictions after the training process. We split our original set in twenty percent for the test set, and for a robust evaluation of network performance, we used $k$-fold cross-validation in training \cite[]{moreno2012study}. We shuffled our validation set in five-folds and then trained the network ten times for each fold using random initial weights. Our training and validation loss always shows the statistics over the fifty training processes. Such strategies account for fluctuations inherent to the random nature of the process.

One crucial aspect to be considered when training a CNN is to define which loss function will be used. The loss is the inaccuracy measure of predictions made by the network and guides the update of the weights through an optimization algorithm. We used the Jaccard loss, especially because of its capability to address class imbalances \cite[]{duque2021power}. Jaccard loss is expressed as:

\begin{equation}
\label{eq:jaccard}
J(y,\hat{y}) = 1 - \frac{ \sum_i (y_{i} \cdot \hat{y}_{i})}{ \sum_i y_{i} + \hat{y}_{i} - (y_{i} \cdot \hat{y}_{i})}
\end{equation}

\begin{figure}
  \centering
  \includegraphics[width=0.45\textwidth]{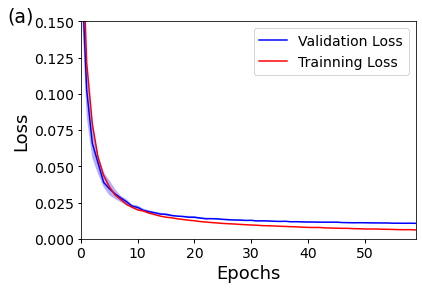}
  \includegraphics[width=0.45\textwidth]{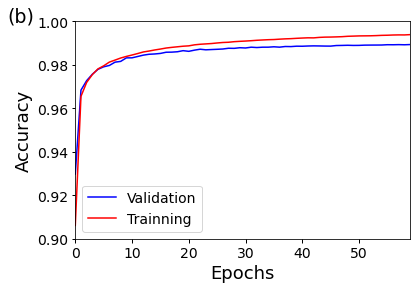}
  \caption{Figure (a) shows the Jaccard loss and (b) shows the Accuracy measured along the training process for the training and validation sets.} 
  \label{fig:trainxval}
\end{figure}

where $y$ is the correct salt mask and $\hat{y}$ is the predicted one. The index $i$ refers to the same pixel in the two images. Equation \ref{eq:jaccard}  evaluates to a minimum of 0.0 when $y$ and $\hat{y}$ intersect entirely, and to a maximum of 1.0 when $y$ and $\hat{y}$ are disjoint. Thus, minimization of the loss represented by Equation \ref{eq:jaccard} fosters the network model to find solutions that increase the intersection between $y$ and $\hat{y}$. Such behaviour explains why the Jaccard loss is ideal for imbalanced problems since it is measured only over the pixels that compose the objects of interest, in this case, the true and predicted salt masks. Nevertheless, we also measure the generalist accuracy score - which is measured for all pixels - as an auxiliary metric.

Figure \ref{fig:trainxval}(a) shows the learning curve corresponding to the network performance and calculated by using the Jaccard loss. The learning curve calculated from the training data set presents how well the network is learning, while the one calculated from the validation data set indicates the generalization capacity of the network. By comparing both curves, it is possible qualitatively infer that the general behavior of the training curves shows an accurate fit, with a slight fractional difference between validation loss and training loss. The validation loss decreases until a point of stability, situated around $50$ epochs, showing a loss median below $0.02$. Figure \ref{fig:trainxval}(b) shows the accuracy, which reaches values up to $0.98$ at the final epochs. This suggests that, despite of the complexity of the problem addressed, the trained network generalizes well.

\section*{Results}

\subsection{Images predicted over the original test set}

\begin{figure}
  \centering
  \includegraphics[scale=0.255,trim={3.5cm 0cm 3.5cm 0cm}]{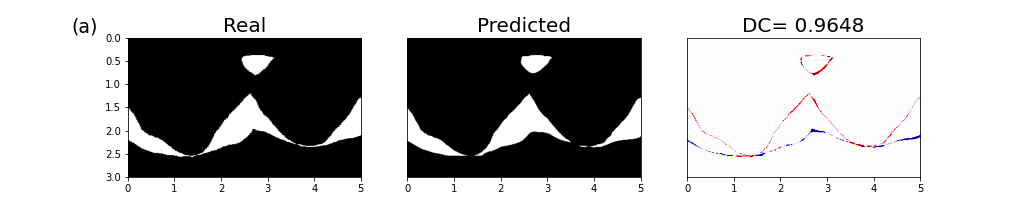}
   \includegraphics[scale=0.255,trim={2.5cm 0cm 3.5cm 0cm}]{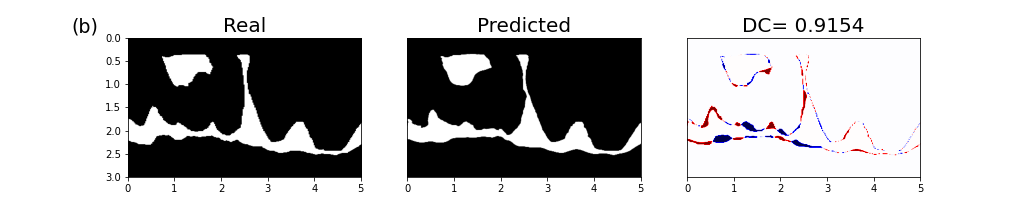} \\
  \includegraphics[scale=0.255,trim={3.5cm 0cm 3.5cm 0cm}]{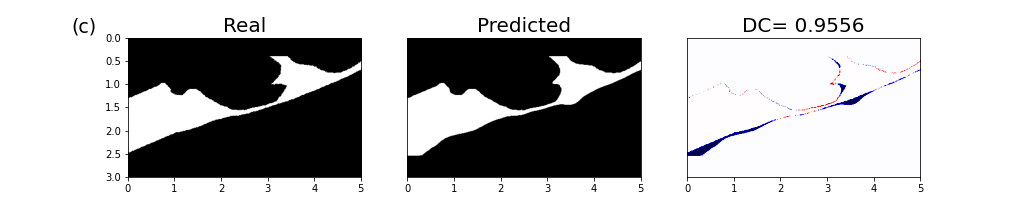}
   \includegraphics[scale=0.255,trim={2.5cm 0cm 3.5cm 0cm}]{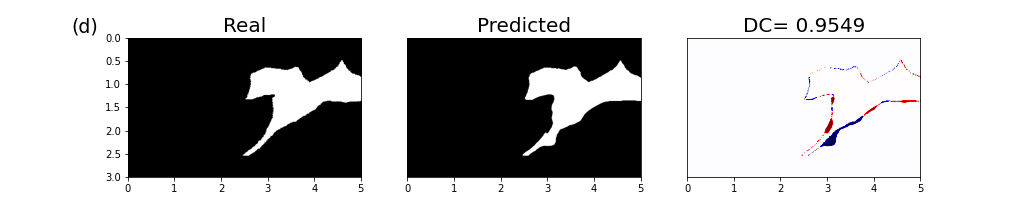} \\ 
     \includegraphics[scale=0.255,trim={3.5cm 0cm 3.5cm 0cm}]{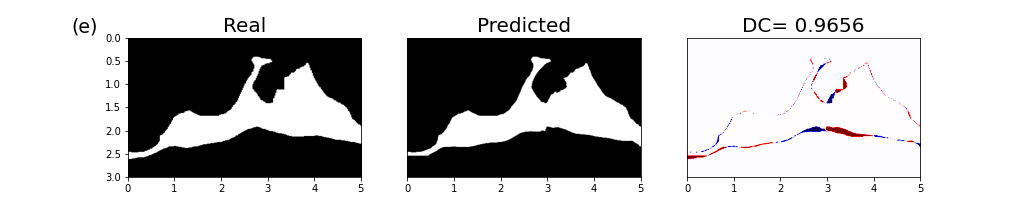}
   \includegraphics[scale=0.255,trim={2.5cm 0cm 3.5cm 0cm}]{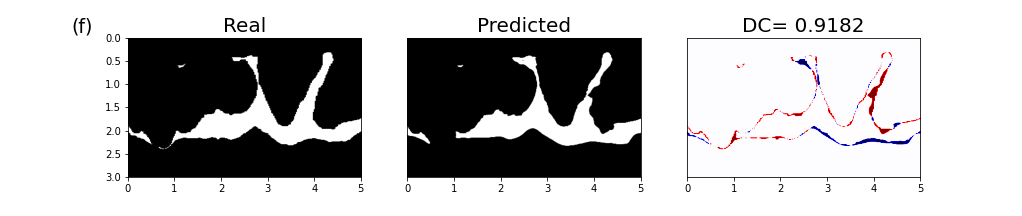} \\
\includegraphics[scale=0.255,trim={3.5cm 0cm 3.5cm 0cm}]{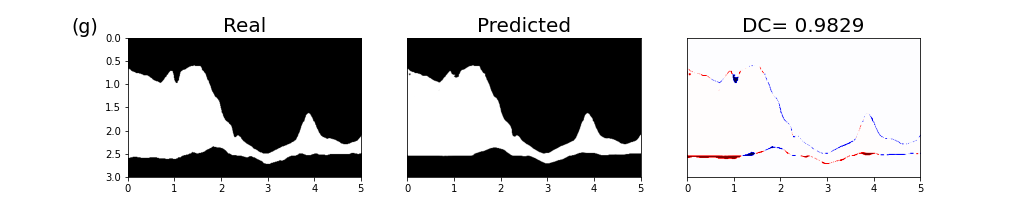}
   \includegraphics[scale=0.255,trim={2.5cm 0cm 3.5cm 0cm}]{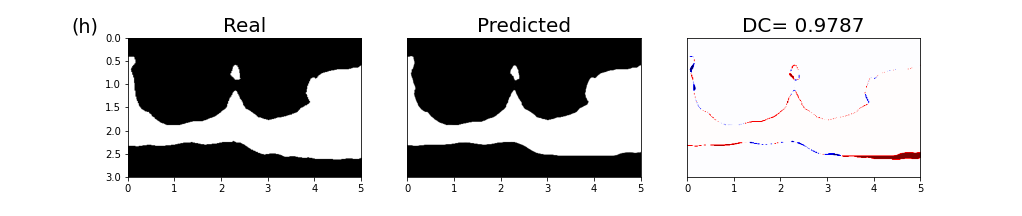} \\ 
     \caption{Examples of masks from the test set where the network made satisfactory predictions, concentrating the errors on the contour details of the salt inclusions. Blue indicates false positive and red false negative in error figures, and DC is the Dice Coefficient measure of each example.}
  \label{fig:pred1}  
\end{figure}

\begin{figure}
  \centering
      \includegraphics[scale=0.255,trim={3.5cm 0cm 3.5cm 0cm}]{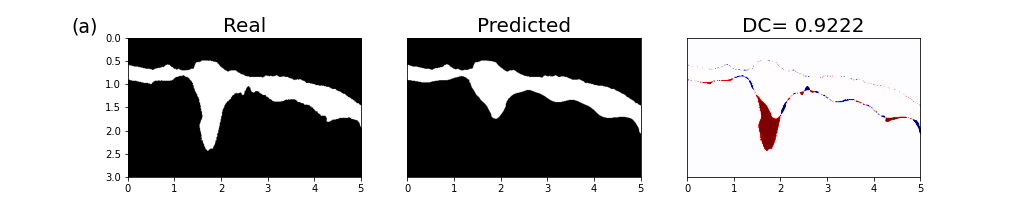}
   \includegraphics[scale=0.255,trim={2.5cm 0cm 3.5cm 0cm}]{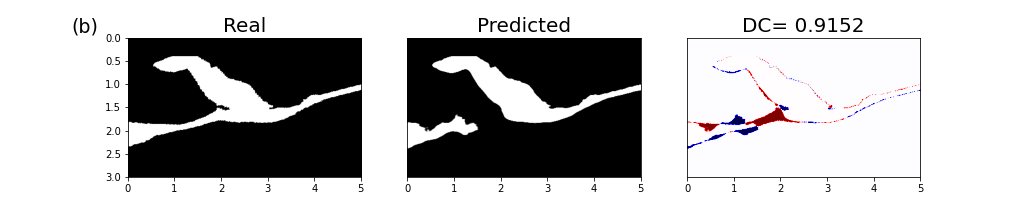} \\ 
   \includegraphics[scale=0.255,trim={3.5cm 0cm 3.5cm 0cm}]{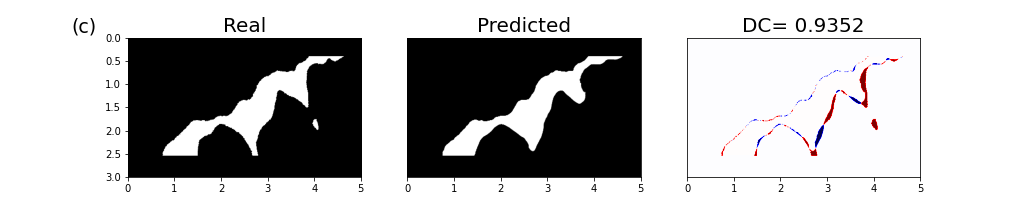}
\includegraphics[scale=0.255,trim={2.5cm 0cm 3.5cm 0cm}]{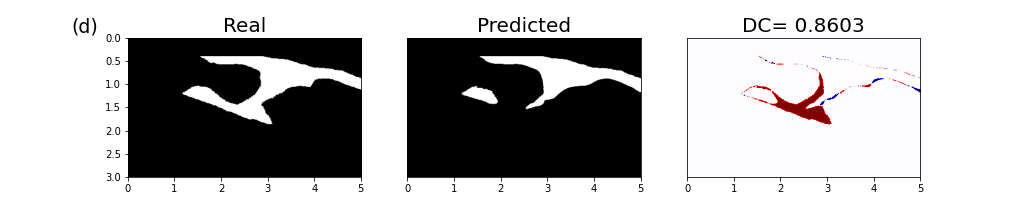} \\
  \includegraphics[scale=0.255,trim={3.5cm 0cm 3.5cm 0cm}]{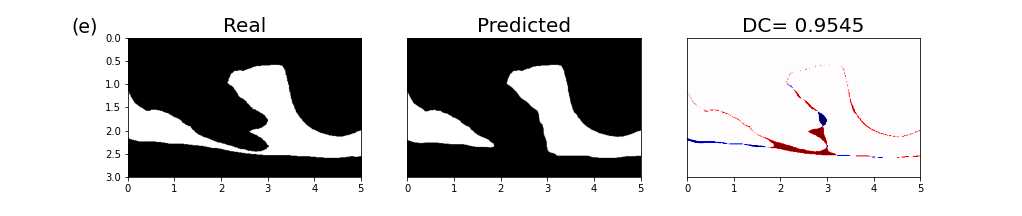}
\includegraphics[scale=0.255,trim={2.5cm 0cm 3.5cm 0cm}]{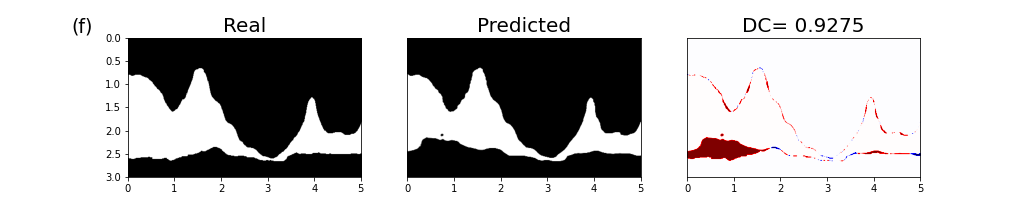} \\

  \caption{Examples of masks from the test set where the network made extensive mistakes, with predictions that do not include some parts of the salt bodies. We consider that these errors are associated with regions with low illumination or the geology is too complex for our simple cross-correlation image condition. Blue indicates false positive and red false negative in error figures, and DC is the Dice Coefficient measure of each example.} 
  \label{fig:pred2}
\end{figure}

We evaluated the results over the test set. The test set is composed of models from the original set that were not seen by the network during training. The results are presented in Figures \ref{fig:pred1} and \ref{fig:pred2}, which show the real salt mask, the predicted salt mask, and the difference between predicted and real masks, where red indicates false-negative and blue indicates false-positive. To measure the accuracy of our predictions on each experiment, the Dice Coefficient \cite[]{dice} is printed above each difference panel in Figures \ref{fig:pred1} and \ref{fig:pred2}. We separated our results into two figures in order to easily discuss the prediction's errors observed.

In Figure \ref{fig:pred1}, we captured models with complex salt structures, such as salt teardrops, overhangs, welds, and tongues, showing that the network can make surprisingly good predictions even for the more challenging cases. One astonishing prevision is the correct shape of allochthonous salt, especially the base of such bodies, and the correct separation from the salt below such structures. The predictions could capture even thin structures, which suggests that this approach can also be helpful to track basalt dikes inside the sediment. The prediction errors made in these masks, even for Figures \ref{fig:pred1}(b) and \ref{fig:pred1}(f), which present the lowest DC, are related to errors in the contour details of the salt inclusions, and could be easily corrected during the FWI process or with interpretation over the stacked migrated image with a velocity model including these predicted salt geometries.

\begin{figure}
  \centering
    \includegraphics[width=0.323\textwidth]{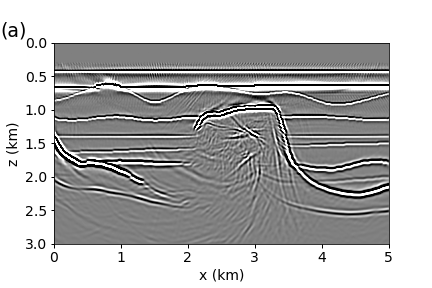}
   \includegraphics[width=0.323\textwidth]{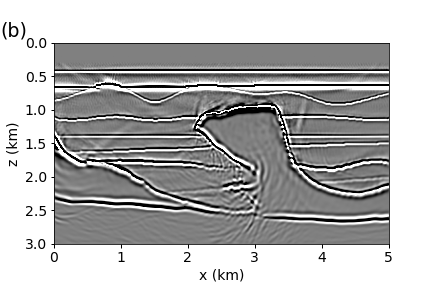} 
   \includegraphics[width=0.323\textwidth]{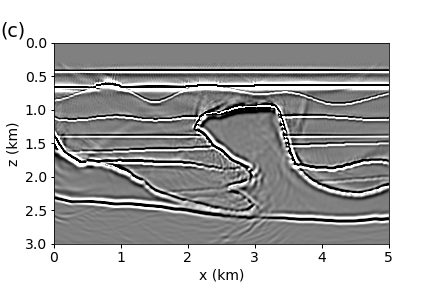} \\
   \includegraphics[width=0.323\textwidth]{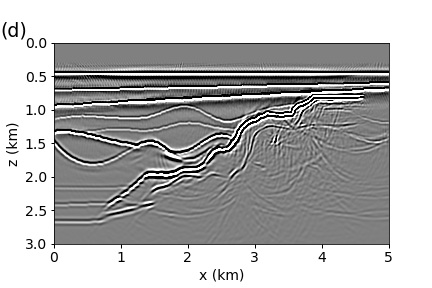}
   \includegraphics[width=0.323\textwidth]{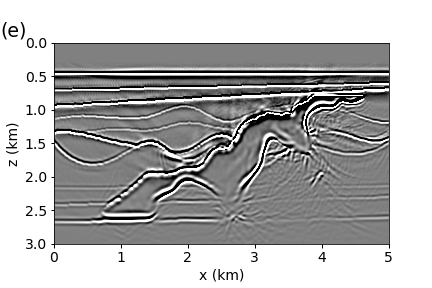} 
   \includegraphics[width=0.323\textwidth]{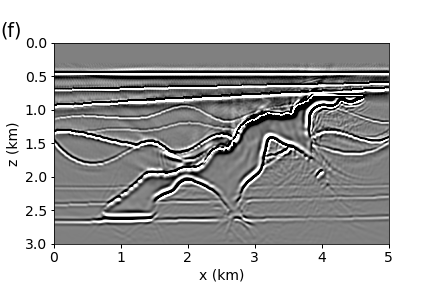} \\
  \caption{Migrated images using the smooth model without salt inclusions ((a) and (d)), the smooth model including the predicted salt mask ((b) and (e)), and the smooth model including the true salt mask ((c) and (f)). The results on (b) use the predicted mask shown in Figure \ref{fig:pred2}(e), and (e) use the predicted mask shown in Figure \ref{fig:pred2}(c).}
  \label{fig:rtmref}
\end{figure}

Although, in Figure \ref{fig:pred2}, one can notice examples where the network made inaccurate predictions, which could jeopardize the migrated image and compromise the interpretation, with some parts of the salt inclusions that were not tracked by the network. Below some overhangs, particularly the thick ones, it is visibly challenging to track the thin salt structures. This is probably due to the lack of illumination necessary to devise such complex structures. We can also note border effects in the bottom and laterals, probably because the energy needed to visualize those structures spread outside the image domains. Another point of attention is the inaccuracy in structures containing salt-tongues or diapirs that are upside down due to the distortion process applied over the geological structures. Our algorithm usually lacks accuracy in these regions, predicting smaller salt than the correct one. Such behavior can be explained by the fact that the bottom of a salt body, when migrated with a smaller velocity, that is our case, collapses in a smaller region than the actual one. Where the base of salt is a thin structure, these collapsed regions are probably smaller than the network can detect as indicative of salt. 

In order to clarify the results that could be obtained when using our salt predictions in a VMB flow, in Figure \ref{fig:rtmref}, we generated images by running RTM using three different models. In the first column, we migrated the synthetic data with the same models used to generate the images for the training/prediction process, i.e., a smooth version of the true models without salt inclusions. Then, in the second column, we included the predicted salt masks over the models used to generate results in the first column. Finally, in the third column, we included the true salt masks over the models used to generate results in the first column. We used the smooth sediment model to emulate the real VMB process, where the true model could not be entirely recovered; thus, we are also considering the effect of velocity errors in the models.

To perform the test presented in Figure \ref{fig:rtmref}, we use the models in Figure \ref{fig:pred2}(c) and (e). They present complex salt structures, like overhangs, whose delineation requires an iterative process when using a conventional VMB flow. Our prediction process makes some mistakes in the salt delineation of these models, as presented in Figure \ref{fig:pred2}(c) and (e). We can see that, despite the misinterpretations in the salt inclusions, the main structures below the salt bodies and the overhang are recovered. Besides, the image generated with the predicted salt can be used to make corrections over the interpreted mask, improving the image quality.
We chose to show the migration without any salt in Figure \ref{fig:rtmref}(a) and (d) only as a reference of the poor quality of images without salt. It is essential to mention that they correspond to the first channel of the input images to the U-Net.

\subsection{Performance evaluation on benchmarks}

\begin{figure}
  \centering
    \includegraphics[width=0.49\textwidth]{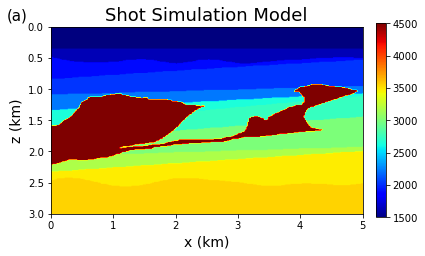}
   \includegraphics[width=0.49\textwidth]{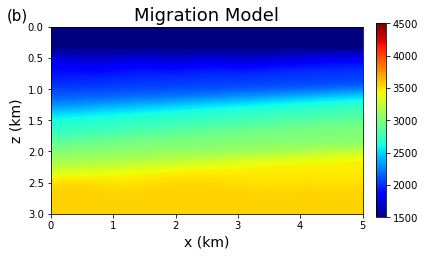} \\
  \caption{Figure (a) shows the velocity model for shot simulation including the salt geometry of a cut in the left portion of the BP/SEG model, and (b) the smoothed sediment model used in migration to generate the input for network predictions.}
  \label{fig:bpmodel}
\end{figure}

\begin{figure}
  \centering
   \includegraphics[scale=0.46,trim={3.5cm 0cm 3.5cm 0cm}]{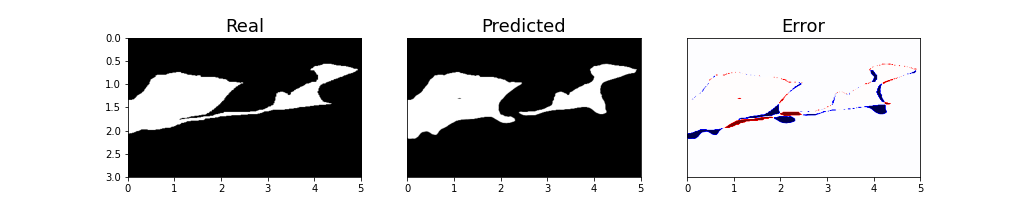} \\
  \caption{Network prediction for the BP/SEG inspired data. Blue indicates false positive and red false negative in error figures.}
  \label{fig:bppred}
\end{figure}

We used two well-known geophysics benchmarks to generate input data sets for our trained network and evaluate its generalization ability. The benchmark data sets were not used during the training/validation process. The first one was a rectangular cutout BP/SEG velocity model \cite[]{BPmodel}. We extracted only the salt masks, re-scaled, and inserted them inside a sediment model as shown in Figure \ref{fig:bpmodel}(a). The sediment model was defined by the same algorithm we used to create our original models. Remember that shapes and velocities of the interfaces are randomly defined but are restricted to a limited range of possibilities. Therefore, using this data, we want to check the generalization of salt geometries, not yet the generality for any sediment velocity. The predicted salt mask for BP/SEG adapted model is shown in Figure \ref{fig:bppred}, where it is possible to see that the network predicted the salt mask with an accuracy of 0.974, restricting the mistakes to regions where there is probably a lack of illumination to precisely devise the reflections.

\begin{figure}
  \centering
      \includegraphics[width=0.49\textwidth]{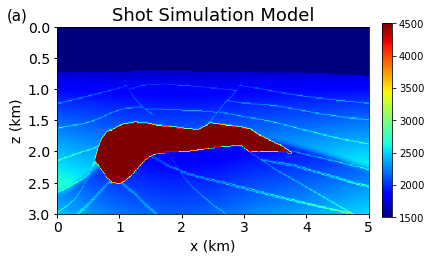}
   \includegraphics[width=0.49\textwidth]{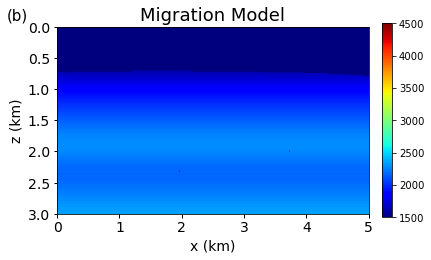} \\
  \caption{Example of one slice of the SEG 3D model used for shot generation (a) and the respective model used for migration (b). One remarkable aspect of these figures is that we do not change sediment's velocity range, which is entirely different from the velocity of our training data set.}
  \label{fig:segmodel}
\end{figure}

Another test was performed over 2D velocity models extracted from the 3D SEG salt model  \cite[]{segsalt2}. In this case, as the SEG model has dimensions more compatible with the ones we used in our work, we only cut and reshaped the model to fit in our interest area, leaving the sediment layers untouched. In Figure \ref{fig:segmodel}, we show one slice of the SEG 3D model used in this work, with the model for shot simulation in Figure \ref{fig:segmodel}(a) and the correspondent model used in RTM in Figure \ref{fig:segmodel}(b). The velocity model used in RTM is a smoothed trace of sediment velocity, which we extended to fulfill our interest area. For this study, we extracted 24 slices from the original 3D model.

\begin{figure}
  \centering
  \includegraphics[scale=0.24,trim={3.5cm 0cm 2.5cm 0cm}]{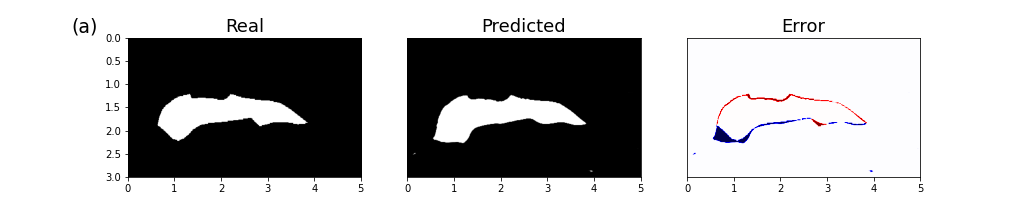}
   \includegraphics[scale=0.24,trim={3.5cm 0cm 2.5cm 0cm}]{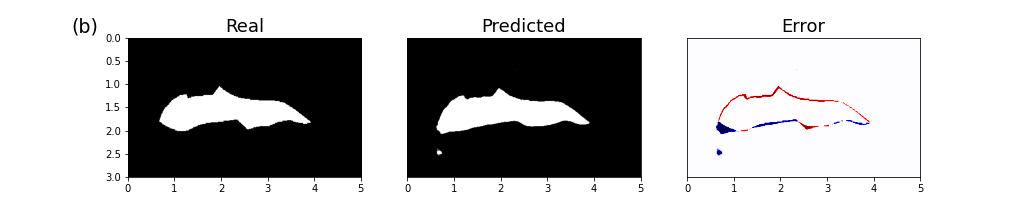} \\
   \includegraphics[scale=0.24,trim={3.5cm 0cm 2.5cm 0cm}]{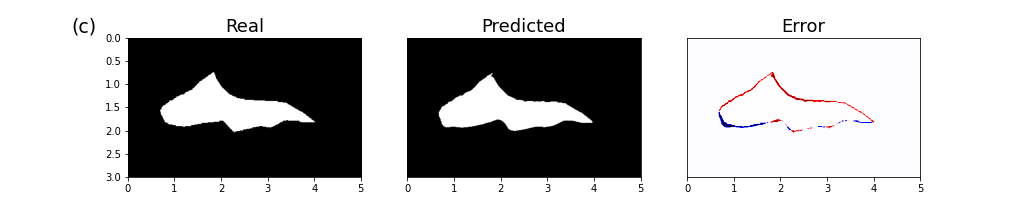}
\includegraphics[scale=0.24,trim={3.5cm 0cm 2.5cm 0cm}]{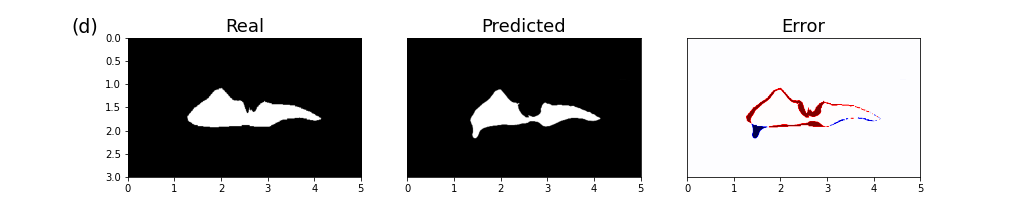}
  \caption{Network predictions for inputs from SEG model. In the error figures, blue indicates false positive and red false negative in error figures.}
  \label{fig:segpred}
\end{figure}

Figure \ref{fig:segpred} presents four examples of predictions made over the SEG data set, where it is possible to see that the network generally made accurate previsions of the salt body location. Here we want to emphasize two differences from the previous predictions we made. First, sediment velocity for the shot generation is entirely different from the one used to generate our training data set. Moreover, the sediment velocity used in migration is slightly worse than the previous case since we used a plane-parallel model in migration due to the lack of an isolated sediment model. The previously mentioned differences in data generated for predictions presented in Figure \ref{fig:segpred} are a strong indication that our trained network is robust enough to generalize the predictions to other interest areas.\\

Another difference between the training set and the SEG test set is that the salt body velocity is 4482~m/s in the SEG models. Initially, we performed a velocity manipulation to set the salt velocity to 4500~m/s, which proved unnecessary with further investigations. The slight deviation from the salt velocity used in training - 4500~m/s to 4482~m/s - did not change the predictions significantly; they are almost equal considering both the DC and the salt body geometry. Extending the tests, we observed that the U-Net preserves the prediction ability for salt velocities between 4350~m/s and 4900~m/s but loses some accuracy in predictions as the velocity deviates too much from the training set velocity of 4500~m/s, especially in the contours of salt inclusions. Despite this encouraging result, we do not believe that pushing the limits of the trained network is the better approach when predicting salt geometries in regions where the salt velocity is too different from 4500~m/s. A better strategy would be to train the U-Net again, using salt models representing such velocity variation.

\subsection{Optimal sampling of the subsurface offsets}

In the initial input data configuration proposed in this work, we use some geophysical premises, such as the sampling of subsurface offset. 
When we build a velocity model using tomography, it is always desirable to have the offset information as denser as possible. Such high density avoids alias effects and makes the assignment of measuring velocity errors easier to be made. In our problem, we do not have to deal with alias effects because tracking of move-out information is not done. Nevertheless, if we consider the information amount, with a large number of offsets, we expect to be giving more information about the salt velocity errors to the network. Thus there are important questions to be answered. For example, which is the minimal offset density necessary for the network to make the correct predictions? Can we ignore kinematics effects and train the network using only one offset?

To verify these points, we changed the network input and measured the performance of the loss during the training process. We defined five different configurations for the input data, as listed below:

\begin{itemize}
\item Dense: $N_{\lambda}$~=~7, with $\lambda$~=~0,20,40,60,80,100,120~m
\item Sparse: $N_{\lambda}$~=~3, with $\lambda$~=~0,60,120~m
\item Near: $N_{\lambda}$~=~1, with $\lambda$~=~0~m
\item Mid: $N_{\lambda}$~=~1, with $\lambda$~=~60~m
\item Far: $N_{\lambda}$~=~1, with $\lambda$~=~120~m
\end{itemize}

We evaluated the performance of validation loss using the different inputs, and observed that the smallest loss is achieved when using the Dense input; the Sparse input has the second-best performance. This result suggests that the network uses all the available information to make the best predictions. When using only one input offset the worst validation loss is achieved, with almost the same values for Near, Mid and Far offset. Such behavior suggests that the network accounts for the kinematics effects to make the salt predictions, being necessary to track the salt reflections positions over the offsets to detect the salt body shape.

\begin{figure}
  \centering
  \includegraphics[width=0.48\textwidth]{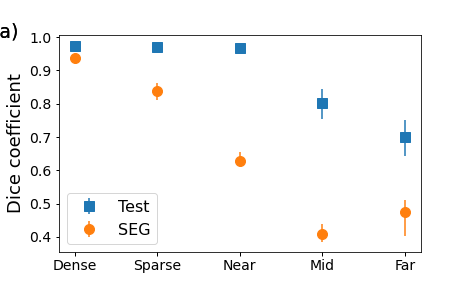}
  \includegraphics[width=0.48\textwidth]{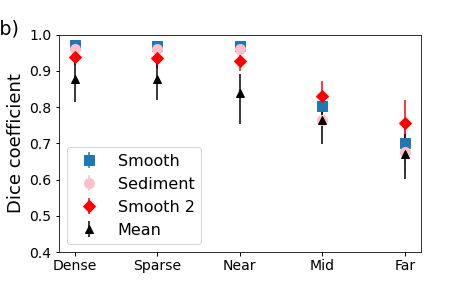}
  \caption{Figure(a) shows the Median Dice Coefficient calculated over each model in the test set and in the SEG models for the different input configurations proposed. Figure(b) shows the median Dice coefficient calculated over the test set when migrated with different sediment velocities. \textit{Smooth} are the true velocities smoothed by a Gaussian with $\sigma = 12$. \textit{Sediment} is the true velocities without salt. \textit{Smooth 2} are the true velocities smoothed by a Gaussian with $\sigma = 40$. \textit{Mean} uses a unique velocity model to migrate all the models, calculated as the mean of all the velocity models from our set.} 
  \label{fig:dice_eval}
\end{figure}

In order to confirm the best way to define the input data, we measured the Dice Coefficient(DC) for each predicted model, 301 in the test set and 24 models in SEG set. Figure \ref{fig:dice_eval}(a) shows the median and one standard deviation as error bars. DC for the test set and SEG models confirms what is observed in the validation loss curve, presenting the best prediction accuracy when using the Dense input. Thus, the results degrade when the number of offsets decreases, reaching the worst results for the Far offset in the test set. 
Therefore, Figure \ref{fig:dice_eval}(a) indicates that our U-Net needs the kinematic information of salt reflections to make accurate predictions. The worst result with Far offset, when we use only one, can be explained by the fact that the top of salt is approximately corrected imaged in the Near offset, due to the correct sediment velocity, and can guide the interpretation of the top of the salt body. 

The effect of the input subsurface offset is more evident when we analyzed DC for SEG models. Figure \ref{fig:dice_eval}(a) shows that the only input configuration that generates reasonable values of DC is the Dense input. With the others inputs, the network is not able to generalize to the SEG model. It is also important to mention here that the maximum subsurface offset of 120~m was chosen based on the analysis of DC over the Test Set. We tested maximum offset ranging from 80~m to 200~m, and DC reaches the highest value with 120~m, not improving when using the larger offsets.

\subsection{How flexible is the trained U-Net?}

Another important question is the influence of the velocity model used to migrate the input data. In Figure \ref{fig:dice_eval}(b), we make a comparison of the prediction performance, measuring the Dice Coefficient when migrating the Test Set using different velocity models. Our original test set is named \textit{Smooth}, migrated with the true sediment model smoothed by a Gaussian filter with $\sigma=12$. \textit{Smooth 2} was migrated with a larger Gaussian filter, with $\sigma=40$. \textit{Sediment} was migrated with the ground truth sediment velocity without the salt inclusion, and \textit{Mean} was migrated using a mean model, which is calculated as the mean of all sediment velocities.

Figure \ref{fig:dice_eval}(b) shows that performance depends on the choice of the velocity models used in migration. The Sediment and Smooth set present almost the same DC, which varies inside the error bars for Dense, Sparse and Near input configurations. Thus, with limited velocities errors like the ones present in the \textit{Smooth} set, which mainly does not affect the travel time in the migration algorithm, or with the True sediment model, the U-Net can recover the salt structures correctly. However, when using velocity models with larger errors, like those in \textit{Smooth 2} and \textit{Mean} sets, the DC reduces, indicating a misinterpretation of the salt. Consequently, Figure \ref{fig:dice_eval}(b) confirms that the conventional practice of inserting the salt body only when the sediment velocity is well resolved is also important for our method.

\begin{figure}
  \centering
  \includegraphics[scale=0.245,trim={2.8cm 0cm 0.5cm 0cm}]{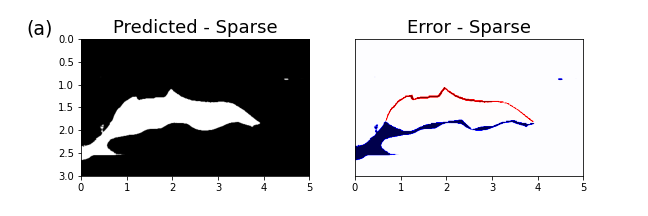}
   \includegraphics[scale=0.245,trim={2.8cm 0cm 0.5cm 0cm}]{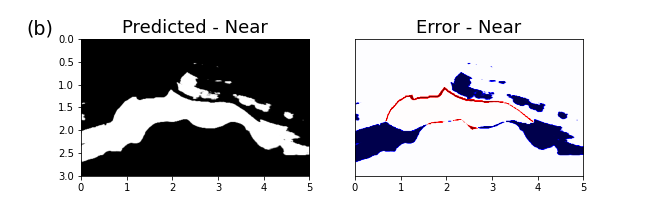}  
   \includegraphics[scale=0.245,trim={2.8cm 0cm 0.5cm 0cm}]{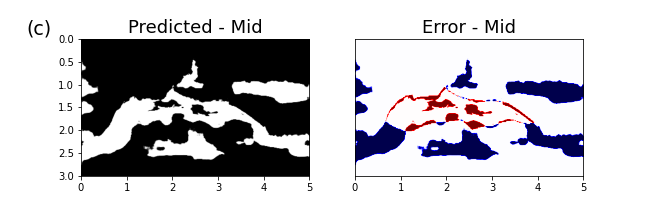}\\
  \includegraphics[scale=0.245,trim={2.8cm 0cm 0.5cm 0cm}]{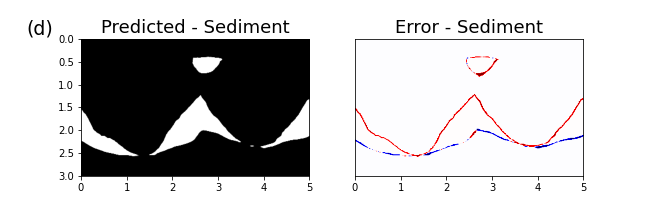}
   \includegraphics[scale=0.245,trim={2.8cm 0cm 0.5cm 0cm}]{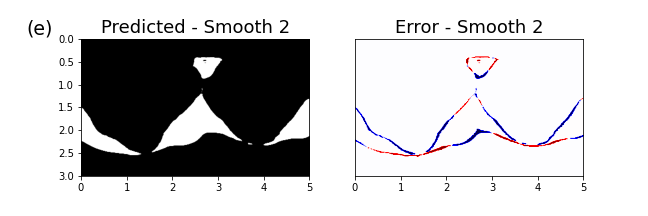} \includegraphics[scale=0.245,trim={2.8cm 0cm 0.5cm 0cm}]{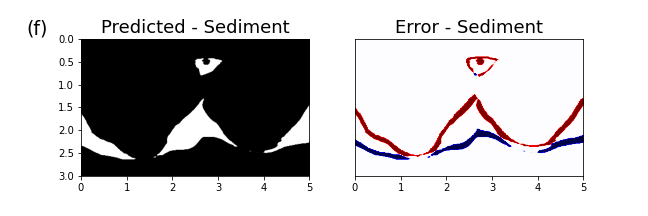}
  \caption{Figures (a-c) shows the predictions when using the Sparse, Near and Mid configurations respectively, made for the salt mask from SEG model in Figure \ref{fig:segpred}(b). Figures (d-f) shows the predictions made with the Dense configuration, but using the Sediment, Smooth2 and Mean models to migrate the input data for the salt mask in Figure \ref{fig:pred1}(a).  Blue indicates false positive and red false negative in error figures.}
  \label{fig:comppred}
\end{figure}

In Figure \ref{fig:comppred} we exemplified the effects observed statistically in Figure \ref{fig:dice_eval}. Figures \ref{fig:comppred}(a-c) show the predictions made over one SEG model (Figure~\ref{fig:segpred}(b)), when using a limited number of input subsurface offsets. It is possible to observe that the predicted images preserve the salt body, but also incorrectly classify sediment areas as salt, confirming that the complete numbers of offsets are necessary to segment the salt from the sediment appropriately. 
In Figures \ref{fig:comppred}(d-f), we show the effect of using different migrations velocities in a Dense configuration, for the model in Figure~\ref{fig:pred1}(a). Checking the predicted images, we observed that with the True sediment model Figure~\ref{fig:comppred}(a), the salt structures were predicted with almost the same quality observed with the Smooth model (Figure~\ref{fig:pred1}(a)). When increasing the error in the model, represented by Figure~\ref{fig:comppred}(b-c), the shape of the salt is almost entirely recovered but presents vertical displacements and lateral misinterpretations. These features can be explained by the fact that errors in the sediment velocities change the position of structures in migrated images. Thus, the U-Net still preserves the ability to identify the salt, but since the sediment velocity model is incorrect, the imaged salt structure presents distortions.

\section{Discussion}

This work proposed a novel use for the subsurface offset, which considers the property that incorrectly migrated reflection spreads through the large subsurface offsets. We trained a U-Net to predict in one step the complete salt inclusion, using as input the shots migrated with a smooth sediment velocity. 
Migrating the shots has two evident improvements over the known strategies of using the shots directly: reduces/regularizes information and brings the network input to the same domain of network output. As we converted the input data to the domain of the output data, this problem can be extended in size or dimensions. However, the migration strategy imposes some requirements not present when using the shots directly, such as the computational cost of migrating the data and the necessity of a well-resolved sediment velocity. Such requirements are new for the VMB flow using DL. However, they are still present in conventional salt inclusion flow, which occurs in a stage when the sediment velocity is well-resolved and requires many migrations until the iterative process of salt inclusion is completed \cite[]{gardenbanks}.

The trained U-Net can partially recover the salt structures using different degrees of smoothness for the sediment velocity, which indicates that the predictions are robust even when we do not know the exact sediment velocity. Besides, migration introduced a generalization ability in the network, as this eliminates the effect of the background velocity. We show with SEG models that we can use a trained network to predict salt even when the sediment velocity is completely different from the one used in the training process, as long as the salt velocity is a value close to the one used for training and the salt geometry in the training set is sufficiently diverse.  
Such generalization is relevant because the most expensive step is generating synthetic data to train the network. Once we have the trained network, predictions are made almost instantly.

When applying an imaging method to reduce our input information and focus only on the desirable feature (salt geometry), we introduced some well-known imaging problems in the predictions, such as illumination, band-limited resolution, and border effects. We chose our survey design, sample interval and wavelet parameters based on previous experience with imaging methods,
reflecting a balanced choice between image quality and computational cost.
Since the quality of predictions are directly related to the quality of the images we input in the U-Net, these questions can be addressed in future investigations of our approach. For example, using seismic data with wide frequency-band, image
condition varying the $\lambda$ horizontally and vertically, time-shift extended imaging condition, and processing improvements to CIGs. Moreover, we believe that an extension to 3D models can better handle the illumination problems that we have in some predictions because the structures could be illuminated in different directions.\\

Applying the salt inclusion technique proposed in this work to real seismic data requires additional studies to deal with some questions. 
Since real acquisitions in complex areas are usually 3D data, one question is the choice of an appropriate CNN architecture to train with 3D images \cite[]{3dunet1,3dunet2,med1,med2} and that can capture the accurate salt inclusions. Since we are training with synthetic data, another question is how to include effects related to noise, anisotropy and multiples in simulations. Another point of attention is that field data could have salt inclusions more complex than the ones we used here, with stratified salt and salt bodies surrounded by complex sediment or high-velocity sediment in deeper areas. Probably the mentioned points should be considered when constructing a training set.
Besides, the phase and amplitude spectrum of the source wavelet in field data vary from the one used in the training process. Thus, it should be necessary to make these wavelets similar before the migration by applying a matching filter \cite[]{paper4}.

\section{Conclusions}

This work presented an alternative approach to the conventional salt flood methodology, which consists of training a DL network (U-Net) to locate salt inclusions over incorrectly migrated subsurface offset panels. The velocity model used in migration must be a good representation of sediment velocities but without any salt inclusion. In migrated images, salt events are entirely distorted from the original geometry due to the lack of salt velocity in the migration model. We showed that a U-Net can successfully learn to identify and make the non-linear transforms necessary to  estimate the salt location when we input a sufficient number of subsurface offsets in the U-Net and a reasonable sediment velocity in migration algorithm. Despite minor inaccuracies, our predicted salt bodies highly correlate with the actual structures. Moreover, when they are included in the migration model, the identification of salt geometry is improved. Thus, even if they can not be used as a final salt interpretation, they can be an excellent initial approximation of the true salt geometry, either for improving the salt body interpretation without the extensive testing of salt scenarios or as a better approximation of salt inclusions for FWI iterations.

\section{Acknowledgements}
APOM thanks Petrobras for sponsoring her postdoctoral research and for the permission to publish this work. 
JCC acknowleges the CNPq financial support through the INCT-GP and the grant 312078/2018-8 and Petrobras. CRB acknowledges the financial support from CNPq (316072/2021-4) and from FAPERJ (grants 201.456/2022 and 210.330/2022). The authors acknowledge the LITCOMP/COTEC/CBPF multi-GPU development team for all the support in the Artificial Intelligence infrastructure and Sci-Mind’s High-Performance multi-GPU system. The authors also would like to thank the SENAI CIMATEC Supercomputing Center for Industrial Innovation, for the cooperation, supply and operation of computing facilities.

\bibliographystyle{seg.bst} 
\bibliography{main}

\newcommand{\SortNoop}[1]{}
\begin{thebibliography}{}
\itemsep0pt

\bibitem[Abadi et~al., 2016]{TF}
Abadi, M., A. Agarwal, P. Barham, E. Brevdo, and Z. Chen,  2016, Tensorflow:
  Large-scale machine learning on heterogeneous distributed systems: arXiv,
  {\bf abs/1603.04467}.

\bibitem[{Alfarhan} et~al., 2020]{SALTUNET}
{Alfarhan}, M., A. {Maalej}, and M. {Deriche},  2020, Concurrent detection of
  salt domes and faults using resnet with u-net: 2020 6th Conference on Data
  Science and Machine Learning Applications (CDMA), 118--122.

\bibitem[Aminzadeh et~al., 1996]{segsalt2}
Aminzadeh, F., N. Burkhard, J. Long, T. Kunz, and P. Duclos,  1996, Three
  dimensional seg/eaeg models — an update: The Leading Edge, {\bf 15},
  131--134.

\bibitem[Araya et~al., 2018]{ARAYAPOLO1}
Araya, M., J. Jennings, A. Adler, and T. Dahlke,  2018, Deep-learning
  tomography: The Leading Edge, {\bf 37}, 58--66.

\bibitem[Araya-Polo et~al., 2019]{ARAYAPOLO2}
Araya-Polo, M., S. Farris, and M. Florez,  2019, Deep learning-driven velocity
  model building workflow: The Leading Edge, {\bf 38}, 872a1--872a9.

\bibitem[Badrinarayanan et~al., 2017]{SEGNET}
Badrinarayanan, V., A. Kendall, and R. Cipolla,  2017, Segnet: A deep
  convolutional encoder-decoder architecture for image segmentation: IEEE
  Transactions on Neural Networks and Learning Systems, {\bf 39}, 2481 -- 2495.

\bibitem[Billette and Brandsberg-Dahl, 2005]{BPmodel}
Billette, F., and S. Brandsberg-Dahl,  2005, The 2004 bp velocity benchmark:
  Presented at the 67th EAGE Conference \& Exhibition, European Association of
  Geoscientists \& Engineers.

\bibitem[Bishop et~al., 1985]{tomo}
Bishop, T.~N., K.~P. Bube, R.~T. Cutler, R.~T. Langan, P.~L. Love, J.~R.
  Resnick, R.~T. Shuey, D.~A. Spindler, and H.~W. Wyld,  1985, Tomographic
  determination of velocity and depth in laterally varying media: Geophysics,
  {\bf 50}, 903--923.

\bibitem[Chen et~al., 2016]{med1}
Chen, J., L. Yang, Y. Zhang, M. Alber, and D.~Z. Chen,  2016, Combining fully
  convolutional and recurrent neural networks for 3d biomedical image
  segmentation: Advances in Neural Information Processing Systems (NIPS), {\bf
  29}, 3036–3044.

\bibitem[{\c{C}}i{\c{c}}ek et~al., 2016]{3dunet1}
{\c{C}}i{\c{c}}ek, {\"O}., A. Abdulkadir, S.~S. Lienkamp, T. Brox, and O.
  Ronneberger,  2016, 3d u-net: Learning dense volumetric segmentation from
  sparse annotation: Medical Image Computing and Computer-Assisted Intervention
  -- MICCAI 2016, Springer International Publishing, 424--432.

\bibitem[Dellinger et~al., 2017]{gardenbanks}
Dellinger, J., A.~J. Brenders, J.~R. Sandschaper, C. Regone, J. Etgen, I.
  Ahmed, and K.~J. Lee,  2017, The garden banks model experience: The Leading
  Edge, {\bf 36}, 151–158.

\bibitem[Dice, 1945]{dice}
Dice, L.~R.,  1945, Measures of the amount of ecologic association between
  species: Ecology, {\bf 26}, 297–302.

\bibitem[Duque-Arias et~al., 2021]{duque2021power}
Duque-Arias, D., S. Velasco-Forero, J.-E. Deschaud, F. Goulette, A. Serna, E.
  Decenci{\`e}re, and B. Marcotegui,  2021, On power jaccard losses for
  semantic segmentation: Presented at the VISAPP 2021: 16th International
  Conference on Computer Vision Theory and Applications.

\bibitem[Esser et~al., 2016]{fwic}
Esser, E., L. Guasch, F. Herrmann, and M. Warner,  2016, Constrained waveform
  inversion for automatic salt flooding: The Leading Edge, {\bf 35}, 235--239.

\bibitem[Etgen et~al., 2009]{etgen2019}
Etgen, J., S.~H. Gray, and Y. Zhang,  2009, An overview of depth imaging in
  exploration geophysics: Geophysics, {\bf 74}, no. 6, WCA5--WCA17.

\bibitem[Frigério and Biondi, 2021]{TWI:IMAGE}
Frigério, O., J., and B. Biondi,  2021, Tomographic waveform inversion (twi):
  Presented at the IMAGE - International Meeting for Applied Geoscience \&
  Energy, SEG |AAPG.

\bibitem[Geng et~al., 2022]{Fomel2021}
Geng, Z., Z. Zhao, Y. Shi, X. Wu, S. Fomel, and M. Sen,  2022, Deep learning
  for velocity model building with common-image gather volumes: Geophysical
  Journal International, {\bf 228}, 1054--1070.

\bibitem[Herron, 2013]{saltscenarios01}
Herron, D.~A.,  2013, Thoughts and observations on interpreting depth-imaged
  data: Interpretation, {\bf 1}, B1--B6.

\bibitem[Hudec and Jackson, 2007]{HUDEC20071}
Hudec, M.~R., and M.~P. Jackson,  2007, Terra infirma: Understanding salt
  tectonics: Earth-Science Reviews, {\bf 82}, 1--28.

\bibitem[Ioffe and Szegedy, 2015]{BNORM}
Ioffe, S., and C. Szegedy,  2015, Batch normalization: Accelerating deep
  network training by reducing internal covariate shift: ArXiv, {\bf
  abs/1502.03167}.

\bibitem[Jones and Davison, 2014]{saltscenarios02}
Jones, I.~F., and I. Davison,  2014, Seismic imaging in and around salt bodies:
  Problems and pitfalls: SEG Technical Program Expanded Abstracts,  3684--3688.

\bibitem[Kazei et~al., 2020]{paper4}
Kazei, V., O. Ovcharenko, and T. Alkhalifah,  2020, Velocity model building by
  deep learning: From general synthetics to field data application: SEG
  Technical Program Expanded Abstracts, 1561--1565.

\bibitem[Kazei et~al., 2021]{ovchar1}
Kazei, V., O. Ovcharenko, P. Plotnitskii, D. Peter, X. Zhang, and T.
  Alkhalifah,  2021, Mapping full seismic waveforms to vertical velocity
  profiles by deep learning: Geophysics, {\bf 86}, no. 5, R711–R721.

\bibitem[Kingma and Ba, 2014]{ADAM}
Kingma, D.~P., and J. Ba,  2014, Adam: A method for stochastic optimization:
  arXiv, {\bf abs/1412.6980}.

\bibitem[Komatitsch and Martin, 2007]{KOMASTISCHPML}
Komatitsch, D., and R. Martin,  2007, An unsplit convolutional perfectly
  matched layer improved at grazing incidence for the seismic wave equation:
  Geophysics, {\bf 72}, no. 5, SM155--SM167.

\bibitem[Lambare et~al., 2014]{RAYTOMO}
Lambare, G., P. Guillaume, and J. Montel,  2014, Recent advances in ray-based
  tomography: European Association of Geoscientists \& Engineers, {\bf 2014},
  1--5.

\bibitem[Lee et~al., 2021]{med2}
Lee, K., L. Sunwoo, T. Kim, and K.~J. Lee,  2021, Spider u-net: Incorporating
  inter-slice connectivity using lstm for 3d blood vessel segmentation: Applied
  Sciences, {\bf 11}.

\bibitem[Li et~al., 2020]{shucai20}
Li, S., B. Liu, Y. Ren, Y. Chen, S. Yang, Y. Wang, and P. Jiang,  2020,
  Deep-learning inversion of seismic data: IEEE Transactions on Geoscience and
  Remote Sensing, {\bf 58}, 2135--2149.

\bibitem[Martin et~al., 2008]{MartinPML}
Martin, R., D. Komatitsch, and A. Ezziani,  2008, An unsplit convolutional
  perfectly matched layer improved at grazing incidence for seismic wave
  propagation in poroelastic media: Geophysics, {\bf 73}, no. 4, T51--T61.

\bibitem[Michell et~al., 2017]{fwisalt01}
Michell, S., X. Shen, A. Brenders, J. Dellinger, I. Ahmed, and K. Fu,  2017,
  Automatic velocity model building with complex salt: Can computers finally do
  an interpreter's job?: SEG Technical Program Expanded Abstracts,  5250--5254.

\bibitem[Moreno-Torres et~al., 2012]{moreno2012study}
Moreno-Torres, J.~G., J.~A. S{\'a}ez, and F. Herrera,  2012, Study on the
  impact of partition-induced dataset shift on $ k $-fold cross-validation:
  IEEE Transactions on Neural Networks and Learning Systems, {\bf 23},
  1304--1312.

\bibitem[Mosher et~al., 2007]{saltflood}
Mosher, C., E. Keskula, J. Malloy, R. Keys, H. Zhang, and S. Jin,  2007,
  Iterative imaging for subsalt interpretation and model building: The Leading
  Edge, {\bf 26}, 1361–1488.

\bibitem[Ovcharenko et~al., 2018]{FWIvariance}
Ovcharenko, O., V. Kazei, D. Peter, and T. Alkhalifah,  2018, Variance-based
  model interpolation for improved full-waveform inversion in the presence of
  salt bodies: Geophysics, {\bf 83}, no. 5, R541–R551.

\bibitem[Pratt, 1999]{pratt}
Pratt, R.~G.,  1999, Seismic waveform inversion in the frequency domain —
  part 1: Theory and verification in a physical scale model: Geophysics, {\bf
  64}, 888–901.

\bibitem[Ronneberger et~al., 2015]{UNET}
Ronneberger, O., P. Fischer, and T. Brox,  2015, U-net: Convolutional networks
  for biomedical image segmentation: Medical Image Computing and
  Computer-Assisted Intervention -- MICCAI 2015, 234--241.

\bibitem[Sava and Vasconcelos, 2011]{extIC}
Sava, P., and I. Vasconcelos,  2011, Extended imaging conditions for
  wave-equation migration: Geophysical Prospecting, {\bf 59}, 35--55.

\bibitem[Sava and Fomel, 2003]{angle}
Sava, P.~C., and S. Fomel,  2003, Angle‐domain common‐image gathers by
  wavefield continuation methods: Geophysics, {\bf 68}, 1065--1074.

\bibitem[Shelhamer et~al., 2016]{LONGSHELHAMER}
Shelhamer, E., J. Long, and T. Darrell,  2016, Fully convolutional networks for
  semantic segmentation: IEEE Transactions on Pattern Analysis and Machine
  Intelligence, {\bf 39}, 1--1.

\bibitem[Shi et~al., 2019]{SALTdetection}
Shi, Y., X. Wu, and S. Fomel,  2019, Saltseg: Automatic 3d salt segmentation
  using a deep convolutional neural network: Interpretation, {\bf 7}, 1A--T725.

\bibitem[Tarantola, 1984]{tarantola}
Tarantola, A.,  1984, Inversion of seismic reflection data in the acoustic
  approximation: Geophysics, {\bf 49}, 1259--1266.

\bibitem[Wang and Nealon, 2019]{DENOISE}
Wang, E., and J. Nealon,  2019, Applying machine learning to 3d seismic image
  denoising and enhancement: Interpretation, {\bf 7}, 1--34.

\bibitem[Wang et~al., 2019]{fwisalt02}
Wang, P., Z. Zhang, J. Mei, F. Lin, and R. Huang,  2019, Full-waveform
  inversion for salt: A coming of age: The Leading Edge, {\bf 38}, 204–213.

\bibitem[Wu et~al., 2019]{3dunet2}
Wu, X., L. Liang, Y. Shi, and S. Fomel,  2019, Faultseg3d: Using synthetic data
  sets to train an end-to-end convolutional neural network for 3d seismic fault
  segmentation: Geophysics, {\bf 84}, no. 3, IM35--IM45.

\bibitem[Yang and Ma, 2019]{YANGBASE}
Yang, F., and J. Ma,  2019, Deep-learning inversion: A next-generation seismic
  velocity model building method: Geophysics, {\bf 84}, no. 4, R583--R599.

\bibitem[Zhang et~al., 2019]{salt:SEG}
Zhang, L., M. Zhang, Z. Zhong, T. Zhao, Y. Wu, J. Wei, and C. Zhan,  2019, Deep
  learning approach in characterizing salt body on seismic images: SEG
  Technical Program Expanded Abstracts, 2594--2598.

\end{thebibliography}

\newpage

\end{document}